\documentclass[reprint, amsmath,amssymb,aps,superscriptaddress,prd,nofootinbib]{revtex4-1}
\usepackage{verbatim}
\usepackage{hyperref}
\hypersetup{colorlinks=true, linkcolor=blue, urlcolor=blue, citecolor=blue}
\usepackage{cleveref}
\usepackage{graphicx}
\usepackage{epstopdf}
\usepackage{dcolumn}
\usepackage{bm}
\usepackage{xspace}
\usepackage[mathlines]{lineno}
\usepackage{subfigure}  
\usepackage{units}
\usepackage{color}
\usepackage{hyperref}
\usepackage{multirow} 
\begin{document}

\title{Search for sterile neutrino mixing using three years \\of IceCube DeepCore data}
\affiliation{III. Physikalisches Institut, RWTH Aachen University, D-52056 Aachen, Germany}
\affiliation{Department of Physics, University of Adelaide, Adelaide, 5005, Australia}
\affiliation{Dept.~of Physics and Astronomy, University of Alaska Anchorage, 3211 Providence Dr., Anchorage, AK 99508, USA}
\affiliation{CTSPS, Clark-Atlanta University, Atlanta, GA 30314, USA}
\affiliation{School of Physics and Center for Relativistic Astrophysics, Georgia Institute of Technology, Atlanta, GA 30332, USA}
\affiliation{Dept.~of Physics, Southern University, Baton Rouge, LA 70813, USA}
\affiliation{Dept.~of Physics, University of California, Berkeley, CA 94720, USA}
\affiliation{Lawrence Berkeley National Laboratory, Berkeley, CA 94720, USA}
\affiliation{Institut f\"ur Physik, Humboldt-Universit\"at zu Berlin, D-12489 Berlin, Germany}
\affiliation{Fakult\"at f\"ur Physik \& Astronomie, Ruhr-Universit\"at Bochum, D-44780 Bochum, Germany}
\affiliation{Physikalisches Institut, Universit\"at Bonn, Nussallee 12, D-53115 Bonn, Germany}
\affiliation{Universit\'e Libre de Bruxelles, Science Faculty CP230, B-1050 Brussels, Belgium}
\affiliation{Vrije Universiteit Brussel (VUB), Dienst ELEM, B-1050 Brussels, Belgium}
\affiliation{Dept.~of Physics, Massachusetts Institute of Technology, Cambridge, MA 02139, USA}
\affiliation{Dept. of Physics and Institute for Global Prominent Research, Chiba University, Chiba 263-8522, Japan}
\affiliation{Dept.~of Physics and Astronomy, University of Canterbury, Private Bag 4800, Christchurch, New Zealand}
\affiliation{Dept.~of Physics, University of Maryland, College Park, MD 20742, USA}
\affiliation{Dept.~of Physics and Center for Cosmology and Astro-Particle Physics, Ohio State University, Columbus, OH 43210, USA}
\affiliation{Dept.~of Astronomy, Ohio State University, Columbus, OH 43210, USA}
\affiliation{Niels Bohr Institute, University of Copenhagen, DK-2100 Copenhagen, Denmark}
\affiliation{Dept.~of Physics, TU Dortmund University, D-44221 Dortmund, Germany}
\affiliation{Dept.~of Physics and Astronomy, Michigan State University, East Lansing, MI 48824, USA}
\affiliation{Dept.~of Physics, University of Alberta, Edmonton, Alberta, Canada T6G 2E1}
\affiliation{Erlangen Centre for Astroparticle Physics, Friedrich-Alexander-Universit\"at Erlangen-N\"urnberg, D-91058 Erlangen, Germany}
\affiliation{D\'epartement de physique nucl\'eaire et corpusculaire, Universit\'e de Gen\`eve, CH-1211 Gen\`eve, Switzerland}
\affiliation{Dept.~of Physics and Astronomy, University of Gent, B-9000 Gent, Belgium}
\affiliation{Dept.~of Physics and Astronomy, University of California, Irvine, CA 92697, USA}
\affiliation{Dept.~of Physics and Astronomy, University of Kansas, Lawrence, KS 66045, USA}
\affiliation{Dept.~of Astronomy, University of Wisconsin, Madison, WI 53706, USA}
\affiliation{Dept.~of Physics and Wisconsin IceCube Particle Astrophysics Center, University of Wisconsin, Madison, WI 53706, USA}
\affiliation{Institute of Physics, University of Mainz, Staudinger Weg 7, D-55099 Mainz, Germany}
\affiliation{Department of Physics, Marquette University, Milwaukee, WI, 53201, USA}
\affiliation{Universit\'e de Mons, 7000 Mons, Belgium}
\affiliation{Physik-department, Technische Universit\"at M\"unchen, D-85748 Garching, Germany}
\affiliation{Institut f\"ur Kernphysik, Westf\"alische Wilhelms-Universit\"at M\"unster, D-48149 M\"unster, Germany}
\affiliation{Bartol Research Institute and Dept.~of Physics and Astronomy, University of Delaware, Newark, DE 19716, USA}
\affiliation{Dept.~of Physics, Yale University, New Haven, CT 06520, USA}
\affiliation{Dept.~of Physics, University of Oxford, 1 Keble Road, Oxford OX1 3NP, UK}
\affiliation{Dept.~of Physics, Drexel University, 3141 Chestnut Street, Philadelphia, PA 19104, USA}
\affiliation{Physics Department, South Dakota School of Mines and Technology, Rapid City, SD 57701, USA}
\affiliation{Dept.~of Physics, University of Wisconsin, River Falls, WI 54022, USA}
\affiliation{Oskar Klein Centre and Dept.~of Physics, Stockholm University, SE-10691 Stockholm, Sweden}
\affiliation{Dept.~of Physics and Astronomy, Stony Brook University, Stony Brook, NY 11794-3800, USA}
\affiliation{Dept.~of Physics, Sungkyunkwan University, Suwon 440-746, Korea}
\affiliation{Dept.~of Physics, University of Toronto, Toronto, Ontario, Canada, M5S 1A7}
\affiliation{Dept.~of Physics and Astronomy, University of Alabama, Tuscaloosa, AL 35487, USA}
\affiliation{Dept.~of Astronomy and Astrophysics, Pennsylvania State University, University Park, PA 16802, USA}
\affiliation{Dept.~of Physics, Pennsylvania State University, University Park, PA 16802, USA}
\affiliation{Dept.~of Physics and Astronomy, University of Rochester, Rochester, NY 14627, USA}
\affiliation{Dept.~of Physics and Astronomy, Uppsala University, Box 516, S-75120 Uppsala, Sweden}
\affiliation{Dept.~of Physics, University of Wuppertal, D-42119 Wuppertal, Germany}
\affiliation{DESY, D-15735 Zeuthen, Germany}

\author{M.~G.~Aartsen}
\affiliation{Department of Physics, University of Adelaide, Adelaide, 5005, Australia}
\author{M.~Ackermann}
\affiliation{DESY, D-15735 Zeuthen, Germany}
\author{J.~Adams}
\affiliation{Dept.~of Physics and Astronomy, University of Canterbury, Private Bag 4800, Christchurch, New Zealand}
\author{J.~A.~Aguilar}
\affiliation{Universit\'e Libre de Bruxelles, Science Faculty CP230, B-1050 Brussels, Belgium}
\author{M.~Ahlers}
\affiliation{Dept.~of Physics and Wisconsin IceCube Particle Astrophysics Center, University of Wisconsin, Madison, WI 53706, USA}
\author{M.~Ahrens}
\affiliation{Oskar Klein Centre and Dept.~of Physics, Stockholm University, SE-10691 Stockholm, Sweden}
\author{I.~Al~Samarai}
\affiliation{D\'epartement de physique nucl\'eaire et corpusculaire, Universit\'e de Gen\`eve, CH-1211 Gen\`eve, Switzerland}
\author{D.~Altmann}
\affiliation{Erlangen Centre for Astroparticle Physics, Friedrich-Alexander-Universit\"at Erlangen-N\"urnberg, D-91058 Erlangen, Germany}
\author{K.~Andeen}
\affiliation{Department of Physics, Marquette University, Milwaukee, WI, 53201, USA}
\author{T.~Anderson}
\affiliation{Dept.~of Physics, Pennsylvania State University, University Park, PA 16802, USA}
% ...

\date{\today}

\collaboration{IceCube Collaboration}
\noaffiliation
\begin{abstract}
We present a search for a light sterile neutrino using three years of atmospheric neutrino data from the DeepCore detector in the energy range of approximately 10--60~GeV.
DeepCore is the low-energy subarray of the IceCube Neutrino Observatory.
The standard three-neutrino paradigm can be probed by adding an additional light ($\Delta m_{41}^2 \sim 1 \mathrm{\ eV^2}$) sterile neutrino. Sterile neutrinos do not interact through the standard weak interaction and, therefore, cannot be directly detected. 
However, their mixing with the three active neutrino states leaves an imprint on the standard atmospheric neutrino oscillations for energies below 100 GeV. 
A search for such mixing via muon neutrino disappearance is presented here. 
The data are found to be consistent with the standard three-neutrino hypothesis. 
Therefore we derive limits on the mixing matrix elements at the level of $|U_{\mu4}|^2 < 0.11 $ and $|U_{\tau4}|^2 < 0.15 $ (90\% C.L.)  for the sterile neutrino mass splitting $\Delta m_{41}^2 = 1.0$ eV$^2$.
\end{abstract}
\interfootnotelinepenalty=10000
\maketitle

\section{Introduction}

Neutrino oscillation is a phenomenon in which a neutrino can be detected as a different weak eigenstate than initially produced after traveling some distance to its detection point.
It arises due to the mixing between neutrino mass and flavor eigenstates and existence of nonzero mass differences between the mass states. The effect is confirmed by a variety of measurements of neutrinos produced in the Sun~\cite{0004-637X-496-1-505,PhysRevC.88.025501,SAGE_PhysRevC.80.015807,GALEX_Hampel1999127,GNO_Altmann2005174,SKsolar_PhysRevD.83.052010}, in the atmosphere~\cite{SKatm_PhysRevLett.81.1562, SKatm_PhysRevLett.93.101801,DeepCore_PhysRevD.91.072004}, at nuclear reactors~\cite{KamLAND_PhysRevLett.100.221803, DayaBay_PhysRevLett.115.111802, DoubleChooz_Abe2014, RENO_PhysRevLett.108.191802}, and at particle accelerators~\cite{MINOS_PhysRevLett.112.191801, T2K_PhysRevLett.112.181801, OPERA_PhysRevD.89.051102,NOvA_PhysRevD.93.051104}. The data from these experiments are often interpreted within the framework of three weakly interacting neutrino flavors, where each is a superposition of three neutrino mass states. However, not all data from neutrino experiments are consistent with this picture. An excess of electron neutrinos in a muon neutrino beam was found at the Liquid Scintillator Neutrino Detector (LSND)~\cite{LSND_PhysRevD.64.112007} and MiniBooNE experiments~\cite{MiniBooNE_PhysRevLett.110.161801}. In addition, the rates of some reactor ~\cite{Reactor_anomaly_PhysRevD.83.073006} and radio-chemical~\cite{Galium_anomaly_PhysRevC.73.045805} experiments are in tension with predictions involving three neutrino mass states. The tension between data and theory can be resolved by adding new families of neutrinos with mass differences \unit[$\Delta m^2 \sim 1 $]{eV$^2$}. %
However, the measurement of the Z$^0$ boson decay width at the Large Electron-Positron (LEP) collider limits the number of the weakly interacting light neutrino states to three~\cite{LEP_PhysRep.427.5-6.257}. This implies that new neutrino species must be ``sterile" and not take part in the standard weak interaction. The simplest sterile neutrino model is a ``3+1" model, which includes three standard weakly interacting (active) neutrino flavors and one heavier\footnote{The effects of the sterile neutrino mixing in the energy range of this study are independent of the sign of $\Delta m_{41}^2$. Therefore the results presented here are also valid for ``1+3", where the sterile state is the lightest. } 
sterile neutrino. The addition of this fourth neutrino mass state modifies the active neutrino oscillation patterns.

The IceCube Neutrino Observatory~\cite{IceCube_detector_Aartsen:2016nxy} is a cubic kilometer Cherenkov neutrino detector located at the geographic South Pole. It is designed to detect high-energy atmospheric and astrophysical neutrinos with an energy threshold of about 100 GeV~\cite{IceCube_HE_nu1_PhysRevLett.111.021103, IceCube_HE_nu2_1242856, IceCube_HE_nu3_PhysRevLett.113.101101, IceCube_HE_nu4_PhysRevD.91.022001, IceCube_HE_nu_global_0004-637X-809-1-98}. DeepCore~\cite{DeepCore_design_Abbasi2012615} is a more densely instrumented subdetector located in the bottom part of the main IceCube array.  
The denser instrumentation lowers the energy detection threshold to \unit[$\sim10$]{GeV}, allowing precision measurements of neutrino oscillation parameters affecting atmospheric muon neutrinos as reported in~\cite{DeepCore_3yrs_disappearance_PhysRevD.91.072004}, where the standard three-neutrino hypothesis is used. This work presents a search for sterile neutrinos within the ``3+1" model framework using three years of the IceCube DeepCore data taken between May 2011 and April 2014. 

An overview of sterile neutrino mixing and its impact on atmospheric neutrino oscillations is presented in Sec.~\ref{sec:mixing} of this article. Section~\ref{sec:icecube} describes the IceCube Neutrino observatory and the DeepCore sub-array used to detect the low energy neutrinos of interest. The selection and reconstruction of atmospheric neutrino events are presented in Sec.~\ref{sec:selection}. A description of the simulation chain, fitting procedure and treatment of systematic uncertainties considered is provided in Sec.~\ref{sec:data_analysis}. Section~\ref{sec:results} presents the results of the search for sterile neutrino mixing. Finally, Sec.~\ref{sec:conclusions} addresses the impact of various assumptions made in the analysis of the data, and places the results of this search into the global picture of sterile neutrino physics. 

\section{Sterile neutrino mixing}
\label{sec:mixing}

The neutrino flavor eigenstates of the weak interaction do not coincide with the mass states, which describe the propagation of neutrinos through space~\cite{BILENKY1978225}. The connection between the bases can be expressed as 
\begin{equation}
\label{mixing3x3}
\left | \nu_\alpha \right > = \sum U^{*}_{\alpha k } \left | \nu_k \right >, 
\end{equation} 
where $|\nu_{\alpha}\rangle$ are the weak states, $|\nu_k\rangle$ are the mass states with mass $m_k$ and $U_{\alpha k}$ are the elements of Pontecorvo--Maki--Nakagawa--Sakata (PMNS) mixing matrix~\cite{BILENKY1978225,Maki01111962} in the standard three-neutrino scenario. For Dirac neutrinos the mixing matrix is parametrized with three mixing angles ($\theta_{12}$, $\theta_{13}$, $\theta_{23}$) and one CP-violating phase. 
Two additional phases are present if neutrinos are Majorana particles, however they play no role in neutrino oscillations.
Muon neutrinos are the main detection channel for DeepCore and are the focus of this study. For the standard three-neutrino model in the energy range of interest for this analysis the muon neutrino survival probability can be approximated as 
\begin{equation}
P (\nu_{\mu} \rightarrow \nu_\mu) \approx 1 -  \sin^2 \left (2 \theta_{23}\right ) \sin^2 \left ( \Delta m_{32}^2 \frac{L}{4 E_{\nu}} \right ),
\label{eq:disappearance}
\end{equation}
where $\Delta m^2_{32} \equiv m^2_{3} - m^2_{2} $ is the mass splitting between states 3 and 2, $\theta_{23}$ is the atmospheric mixing angle, $L$ is the distance traveled from the production point in the atmosphere and $E_{\nu}$  is the neutrino energy. The diameter of the Earth and size of the atmosphere define the baselines that range between 20 and 12700 km. 

The addition of a single sterile neutrino, $\nu_s$, with corresponding mass eigenstate denoted as $\nu_4$, modifies the mixing matrix in Eq. (\ref{mixing3x3}) as  
\begin{equation}
\mathbf{U} \equiv 
\left ( 
\begin{matrix}
U_{e 1} && U_{e 2} && U_{e 3} && U_{e 4} \\
U_{\mu 1} && U_{\mu 2} && U_{\mu 3} && U_{\mu 4} \\
U_{\tau 1} && U_{\tau  2} && U_{\tau 3} && U_{\tau 4} \\
U_{s 1} && U_{s 2} && U_{s 3} && U_{s 4} 
\end{matrix}
\right ). 
\end{equation}
A single sterile neutrino family adds six new parameters~\cite{Kopp2013}: three mixing angles $\theta_{14}$, $\theta_{24}$, $\theta_{34}$,  two CP-violating phases $\delta_{14}$, $\delta_{34}$ and one mass difference $\Delta m_{41}^2$. 
IceCube has no sensitivity to CP-violating phases and, therefore, they are assumed absent in this study. 
In this case the 4$\times$4 mixing matrix can be parametrized~\cite{Kopp2013} as 
\begin{equation}
\mathbf{U} = \mathbf{U}_{34} \mathbf{U}_{24} \mathbf{U}_{23} \mathbf{U}_{14} \mathbf{U}_{13} \mathbf{U}_{12},\label{mixing4x4}
\end{equation}
where $\mathbf{U}_{ij}$ is a rotation matrix by an angle $\theta_{ij}$ in the $ij$-plane.
\begin{figure}[!t]
  \centering
    \includegraphics[width=0.5\textwidth]{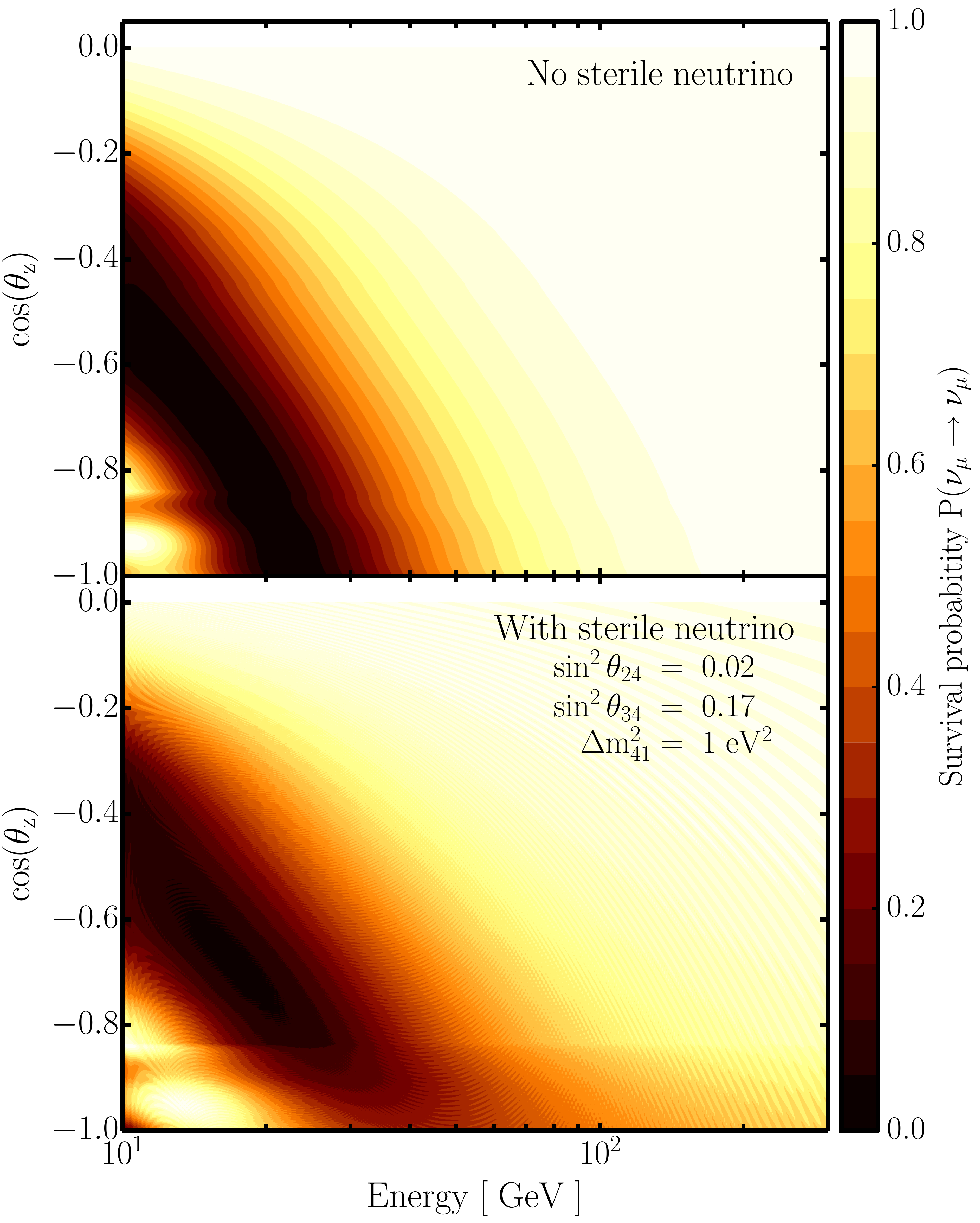}
  \caption{The muon neutrino survival probability for (top) the standard three-neutrino oscillations and (bottom) ``3+1'' sterile neutrino model as function of true muon neutrino energy and the cosine of the true neutrino zenith angle $\theta_{z}$.  Values $\Delta m_{32}^2 = 2.51 \cdot 10^{-3} $ eV$^2$, $\sin^2 \theta_{23} = 0.51$ are assumed for the standard atmospheric mixing parameters. }
  \label{oscillograms}
\end{figure}

The mixing angle $\theta_{14}$ affects mainly electron neutrinos, which have only a minor impact on this study. Therefore the mixing matrix can be simplified further by setting $\theta_{14}$ to zero. 
These assumptions simplify the elements of $\mathbf{U}$ describing the mixing of the active states to the sterile neutrino state~\cite{Razzaque2011}:
\begin{equation}
\begin{split}
\left |U_{e4} \right |^2 &= 0 , 
\\ 
\left |U_{\mu 4} \right |^2 &= \sin^2 \theta_{24}, 
\\
\left |U_{\tau 4} \right |^2 &= \cos^2 \theta_{24} \cdot \sin^2 \theta_{34} . 
\end{split}
\end{equation}
This additional sterile neutrino state modifies the muon neutrino oscillation pattern~\cite{PhysRevD.85.093010,Nunokawa2003279}. 

The propagation of neutrinos is described by the Schr\"{o}dinger equation 
\begin{equation}
\label{schroedinger}
i \dfrac{d}{dx} \Psi_{\alpha} = \hat{H}_F \Psi_{\alpha} ,
\end{equation}
where $x$ is a position along the neutrino trajectory, $\Psi_{\alpha}=(\nu_{e}, \nu_{\mu}, \nu_{\tau}, \nu_s )^{T}$, and $\hat{H}_F$ is an effective Hamiltonian 
\begin{equation}
\label{hamiltionian}
\hat{H}_F = \frac{1}{2 E_{\nu}}\mathbf{U} \hat{M}^2 \mathbf{U^\dag} + \hat{V}_{int},
\end{equation}
where  $\mathbf{U}$ is the mixing matrix described in Eq. (\ref{mixing4x4}), $\hat{M^2}$ is the neutrino mass matrix, and $\hat{V}_{int}$ is an interaction potential. For neutrinos passing though neutral matter, the interaction part of the Hamiltonian in Eq. (\ref{hamiltionian}) can be expressed as 
\begin{equation}
\label{potential}
\hat{V}_{int} \equiv \pm \frac{G_F}{\sqrt{2}} \operatorname{diag}(2 N_e, 0, 0, N_n), 
\end{equation}
where the sign $+(-)$ corresponds to neutrinos (antineutrinos), $G_F$ is Fermi's constant, and $N_e$ and $N_n$ are the densities of the electrons and the neutrons in matter, respectively.

All active neutrinos have a matter potential due to weak neutral current (NC) interaction while sterile neutrinos do not interact with matter at all. This can be expressed as an effective matter potential for the sterile neutrino states equal to the matter potential of NC interactions for active neutrinos with an opposite sign.

The probability of a $\nu_{\alpha}$ to $\nu_{\beta}$ transition is calculated as  
\begin{equation}
\label{probability}
P_{\alpha\beta} = P(\nu_\alpha \to \nu_\beta) = \left | \langle \nu_{\beta}  \vert \nu_{\alpha}(x) \rangle \right | ^2,
\end{equation}
where $\nu_{\alpha}(x)$ is a solution of Eq. (\ref{schroedinger}). It is nontrivial to solve Eq. (\ref{schroedinger}) analytically for atmospheric neutrinos crossing the Earth. Therefore, the probabilities are calculated numerically including all mixing parameters in a ``3+1" model using the 12-layer approximation of the Preliminary Reference Earth Model (PREM)~\cite{PREM_DZIEWONSKI1981297} and the General Long Baseline Experiment Simulator (GLoBES)~\cite{GLoBES_Huber:2004ka, GLoBES_Huber:2007ji}.

The upper panel of  Fig.~\ref{oscillograms} shows the survival probability  for atmospheric muon neutrinos as a function of true energy and zenith angle, $\theta_{z}$,
in the case of the standard three-neutrino oscillations. For the neutrinos crossing the Earth by the diametral trajectory ($\cos \theta_z = -1$) the minimum survival probability is at approximately 25 GeV. The atmospheric neutrino mixing is close to maximal ($\theta_{23} \sim 45^\circ$), which leads to almost complete disappearance of muon neutrinos.
The minimum of the oscillation pattern follows Eq. (\ref{eq:disappearance}) and does not change its depth or show discontinuities between different arrival directions.

The addition of a sterile neutrino state modifies the neutrino oscillations in two ways that are relevant for this analysis. The first is connected to vacuum oscillations into the sterile neutrino state. 
These fast oscillations cannot be resolved at the final analysis level and instead result in a change of the overall flux normalization. The second effect is caused by the different effective matter potential experienced by the sterile neutrino state when crossing the Earth. This modifies the amplitude and energy of the muon neutrino oscillation minimum. 
The strength of the change is proportional to the amount of matter along the neutrino trajectory, and is, therefore, more pronounced for neutrinos crossing the Earth's core. This is demonstrated in the bottom panel of Fig.~\ref{oscillograms}, where the largest change in the muon neutrino survival probability is seen for trajectories with $\cos \theta_z < -0.8$. 

\begin{figure}[!t]
  \centering
    \includegraphics[width=0.5\textwidth]{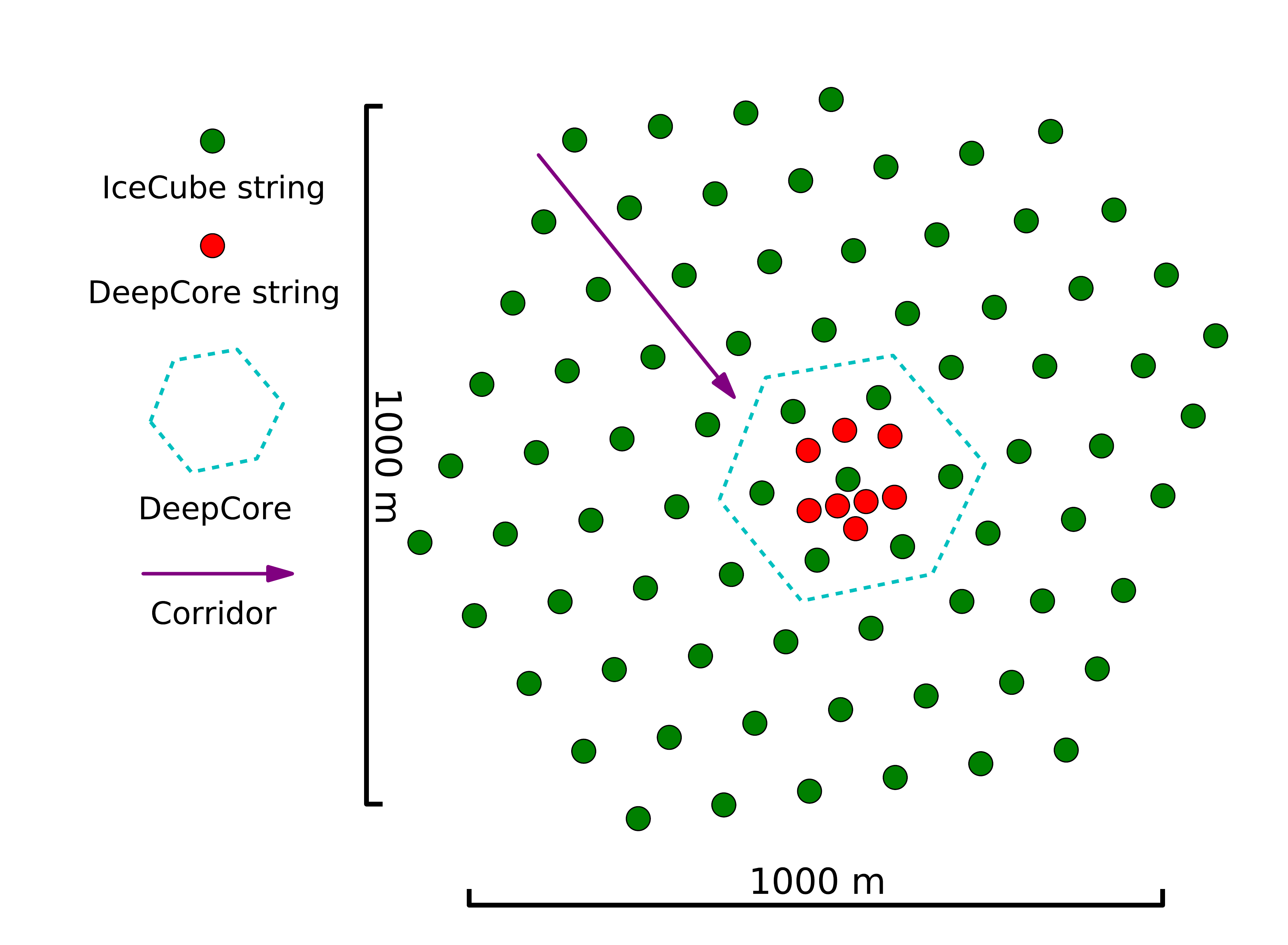}
  \caption{The top view of IceCube. Green circles indicate positions of the ordinary IceCube strings. Red circles show the configuration of the DeepCore strings with denser instrumentation and high quantum efficiency DOMs. The dashed line encompasses the DeepCore area of the detector. The purple arrow shows an example of the corridor direction formed by the detector geometry. }
  \label{DeepCore_topview}
\end{figure}

The value of the sterile mass splitting $\Delta m^2_{41}$ changes only the period of oscillations between muon and sterile states. 
Such oscillations are averaged by the detector energy and zenith resolutions and cannot be resolved for neutrinos with energies considered in this study. 
Therefore, throughout this analysis $\Delta m^2_{41}$ is fixed to \unit[1]{ eV$^2$}. The impact of these assumptions is discussed in Sec.~\ref{sec:conclusions}.  

The light (standard) neutrino mass ordering influences the effects of the sterile neutrino mixing. Switching from one assumed mass ordering to the other interchanges the oscillation probabilities for neutrinos and antineutrinos~\cite{Nunokawa2003279}. This effectively leads to some sensitivity to the standard neutrino mass ordering if both mixing elements $|U_{\mu4}|^2$ and $|U_{\tau4}|^2$ are significantly nonzero~\cite{PhysRevD.85.093010}.

\begin{figure}[!t]
  \centering
    \includegraphics[width=0.5\textwidth]{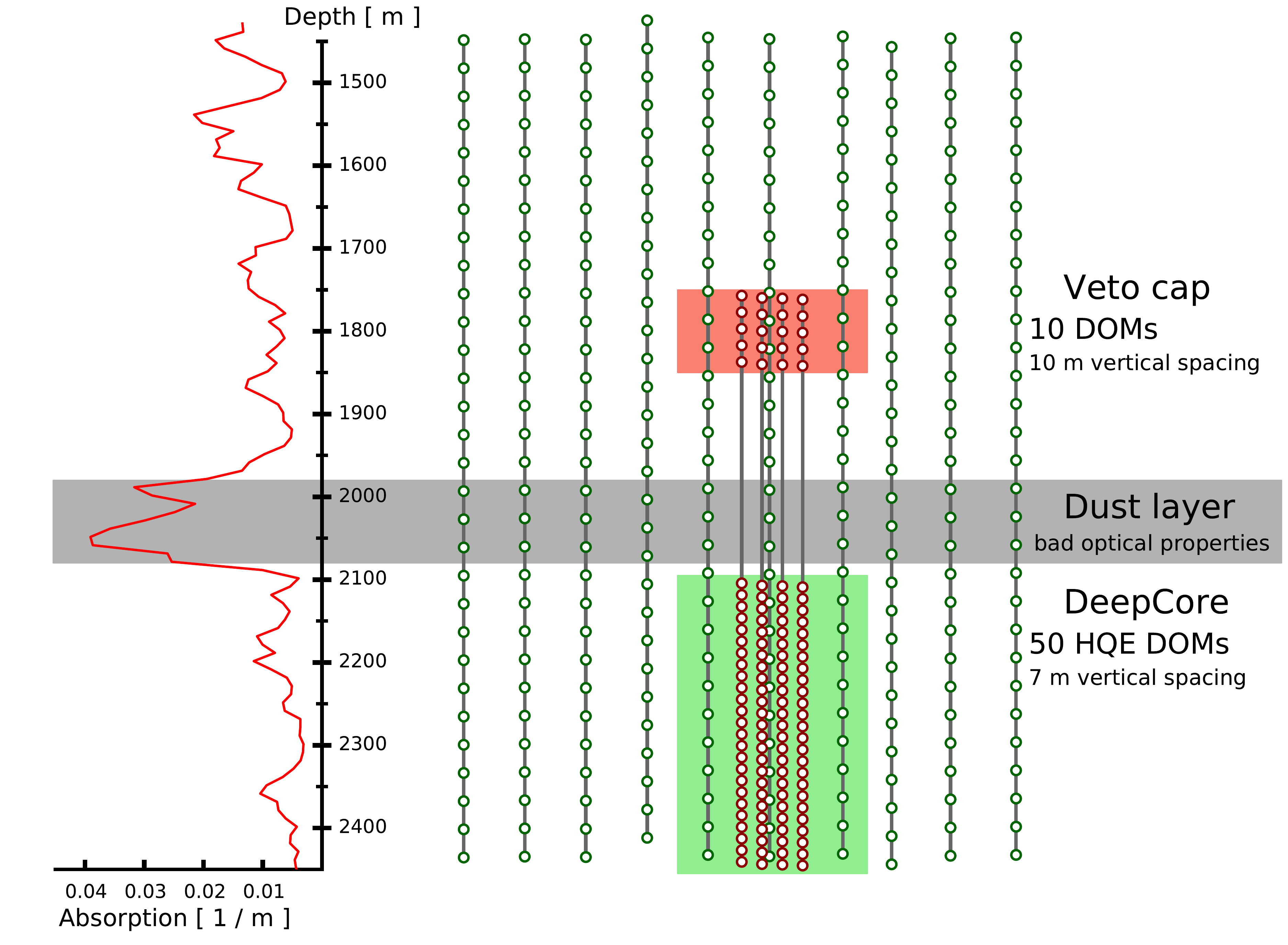}
  \caption{The side view of the IceCube experiment. Green and red circles represent the standard IceCube DOMs and high quantum efficiency DeepCore DOMs, respectively. The dust layer, a region with short optical absorption length, is highlighted gray. The green region shows the DeepCore fiducial volume, and the red region is used to improve the veto efficiency against down-going atmospheric muons. The red line on the left axis shows the optical absorption length as function of depth for the optical ice model used in the study~\cite{SpiceLea_ICRC2013}. }
  \label{DeepCore_sideview}
\end{figure}

At higher energies, muon anti-neutrinos can undergo resonantlike transitions~\cite{Petcov:2016iiu} to the sterile state.
This happens when the neutrino energy, sterile mixing and mass splitting meet the criteria for the mantle--core parametric enhancement~\cite{Chizhov:1998ug,NOLR_PhysRevD.63.073003} due to matter effects~\cite{Wolfenstein_PhysRevD.17.2369,Mikheev:1986gs} in Earth. 
The resonant transition results in a deficit of muon antineutrinos compared to the expectation from the standard neutrino mixing for neutrinos with energies above 1 TeV that cross the Earth's core. 
A search for such a transition has been published by IceCube~\cite{IC86_HEsteriles_PhysRevLett.117.071801}. Since this effect is pronounced at energies above 1 TeV it has no impact on this study.

\section{IceCube DeepCore detector}
\label{sec:icecube}

The IceCube neutrino detector uses the antarctic ice as a natural optical medium to detect the Cherenkov light from secondary particles produced in neutrino interactions in or near the detector. The detector instruments about 1 km$^3$ of ice with digital optical modules (DOMs) arranged in an array of 86 strings with 60 modules each~\cite{DOM_Abbasi2009294,DOM_Abbasi2010139}. The strings are arranged in a hexagonal grid with typical inter-string separation of 125~m, except for the 8 DeepCore strings, which are placed closer together in the center of the array at a typical distance of 50~m. The vertical DOM separation is 17~m, except in the DeepCore strings, where it is 7 m. Each DOM contains a downward-looking 10" photomultiplier tube and digitizing electronics enclosed in a pressure resistant glass sphere. The DOMs are located at depths between 1450~m and 2450~m below the ice surface. 

The DOMs composing the DeepCore strings are equipped with 35\% higher quantum efficiency photomultiplier tubes to increase light collection. The reduced spacing between DeepCore modules lowers the energy threshold of the detector to about 10 GeV. A top and side view of the DeepCore position inside IceCube are shown in Fig.~\ref{DeepCore_topview} and~\ref{DeepCore_sideview}, respectively.  This study uses the 8 DeepCore strings along with the surrounding IceCube strings as a definition of the DeepCore detector as denoted in Fig.~\ref{DeepCore_topview}.  

The remaining outer layers of the IceCube array are used as a veto-detector against the prevailing background from atmospheric muons. IceCube DeepCore has a baseline of up to \unit[12700]{km}, depending on the neutrino arrival direction. 
This, together with the low energy threshold and a large instrumented volume, makes the DeepCore detector a unique tool in the study of atmospheric neutrino oscillations.

\section{Event selection and reconstruction}
\label{sec:selection}

The event selection in this analysis aims to identify charged current (CC) muon (anti)neutrino events with interaction vertices contained within the DeepCore detector volume. A muon track and a hadronic shower are produced in CC interactions. 
The selection is also designed to reduce the large background contribution from atmospheric muons produced in cosmic ray interactions. Details of the event selection are outlined in~\cite{DeepCore_3yrs_disappearance_PhysRevD.91.072004} and~\cite{Yanez_thesis}. Here we review the key components of the selection.

\subsection{Background rejection}

The first step in the event selection involves a dedicated DeepCore trigger and data filter that is designed to select neutrino-induced events and reject atmospheric muon events~\cite{DeepCore_design_Abbasi2012615}.  
The events reconstructed as down-going ($\cos \theta_z > 0$) by a fast track reconstruction algorithm~\cite{LineFit_Aartsen2014143} or a maximum likelihood reconstruction~\cite{SPEFit_Ahrens2004169} are rejected. 
A small fraction of down-going atmospheric muons can be misreconstructed as up-going. However, due to the large atmospheric muon flux, this small fraction can still lead to a large contamination in the final data sample. 

Additional algorithms are used to identify and reject the remaining atmospheric muon background. 
The position of the earliest DOM triggering the detector is required to be inside the DeepCore volume. 
This requirement selects up-going events starting inside the DeepCore volume, but rejects down-going atmospheric muons, which have to pass through the outer IceCube strings and, therefore, leave the first signals there. 
In addition, background events are identified using the observed charge in the upper part of IceCube, accumulated charge as a function of time ($dQ/dt$) and charge observed before the trigger~\cite{Yanez_thesis}.

The most powerful veto criterion against remaining atmospheric muons is the \textit{corridor cut}. This algorithm identifies muons that penetrate the detector through the corridors formed by the geometry of the detector configuration.
This cut rejects events if two or more DOMs register a signal within a narrow time window [--150 ns, +250 ns] from the expected arrival time of Cherenkov light coming from an atmospheric muon traveling through a corridor. An example of such a direction is depicted in Fig.~\ref{DeepCore_topview}.
A requirement of more than two hits in the corridor veto region is used to select a data driven sample of atmospheric muons and to construct a background template. 

\begin{figure}[!t]
  \centering
    \includegraphics[width=0.35\textwidth]{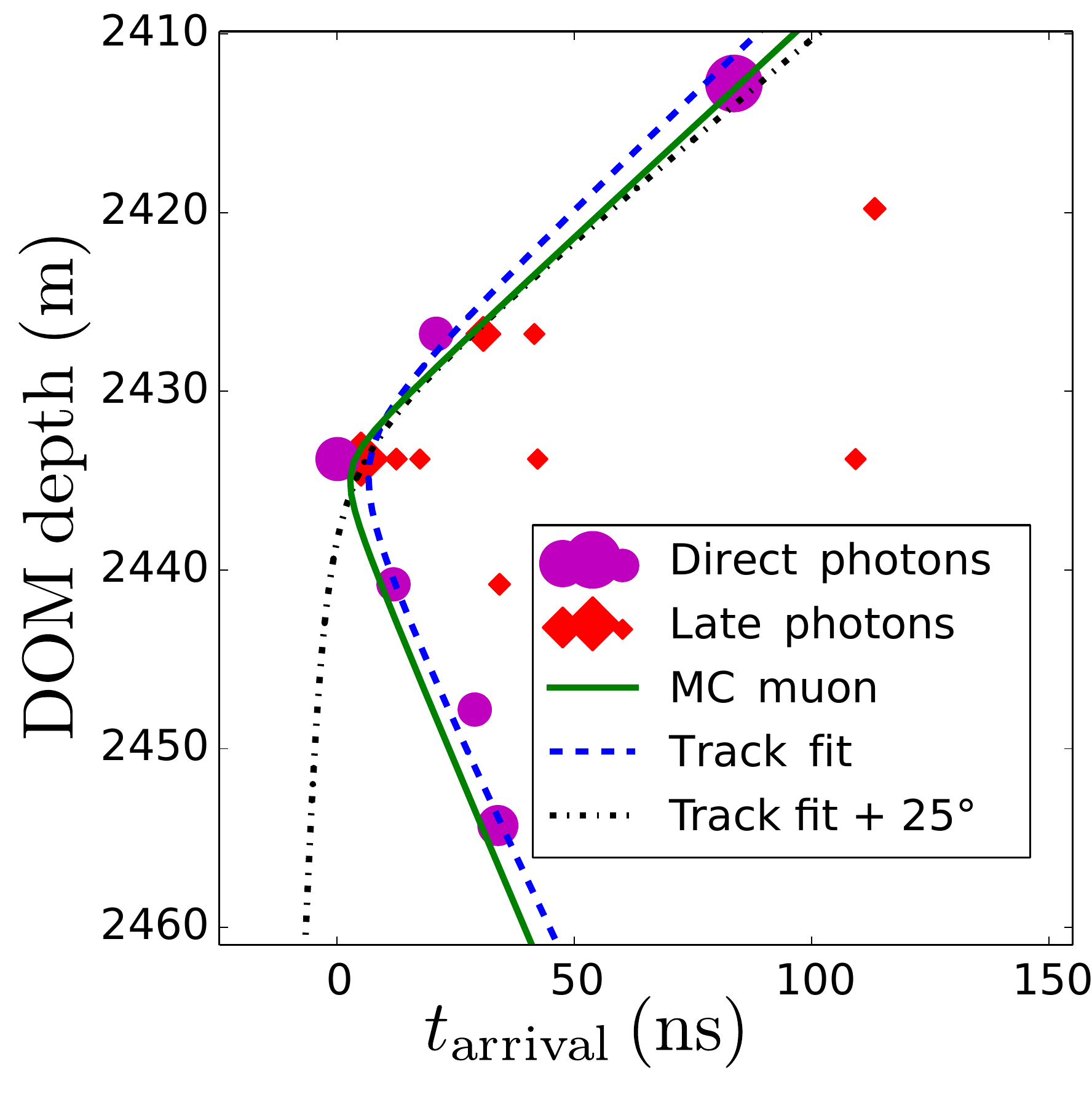}
  \caption{A hyperbolic light pattern in time and DOMs depth created by the direct photons from a muon track passing next to a string. Magenta and red markers depict direct and scattered (late) photons, respectively. The solid green line shows the expectation from the true muon.
The dashed blue curve depicts the fitted hyperbola of the reconstructed muon track and dot-dashed black curve shows the expectation if the direction is changed by 25$^\circ$~\cite{DeepCore_PhysRevD.91.072004}.}
  \label{fig:hyperbola}
\end{figure}

The criterion on the position of the first DOM triggered in the event is strengthened as compared to~\cite{Yanez_thesis}. In this study it is required to be in the bottom \unit[250]{m} of the detector. This provides a buffer zone between the acceptable DeepCore fiducial volume and the ``dust layer'' shown as gray in Fig.~\ref{DeepCore_sideview}. This region, characterized by a short optical absorption length, is present due to dust accumulation during a geological period about 60 to 70 thousand years ago~\cite{DustLogger:2013:0022-1430:1117}.  Atmospheric muons that enter the detector through the dust layer leave few traces to satisfy veto criteria and can mimic up-going neutrinos. The addition of a buffer layer reduces contamination from such events.

\subsection{Reconstruction of $\nu_\mu$ interactions}
Near the detector energy threshold, neutrino interactions are likely to be detected only if  they happen near a detector string. These events will leave signals in only a few DOMs.
Most of the Cherenkov photons undergo scattering, but using \textit{direct} (i.e. nonscattered) photons minimizes the impact of uncertainties of the optical properties of the ice.

The selection of direct photons uses the fact that the Cherenkov light is emitted at a characteristic angle relative to the direction of the muon produced in the $\nu_{\mu}$ CC interaction. Therefore, the depth at which nonscattered photons arrive at DOMs on a string is a hyperbolic function of time~\cite{Aguilar2011652} as shown in Fig.~\ref{fig:hyperbola}. Scattered or \textit{late} photons have an additional time delay and do not match the hyperbolic pattern. A time window for accepting direct photons is defined based on the vertical distance between two DOMs and the time it would take nonscattered photons to travel such a distance in ice. A time delay up to 20 ns is allowed in this analysis. 
Signals from at least three triggered DOMs are required to meet this direct photon selection criteria.

The direct photons of an event are used to fit tracklike (muon) and pointlike (hadronic or electromagnetic shower) emission patterns of Cherenkov light using a $\chi^2$ optimization. The ratio of the $\chi^2$ values for the two hypotheses is used to select tracklike events, which are likely to be caused by $\nu_\mu$ CC interactions. This selection rejects about 35\% of all $\nu_\mu$ CC interactions. 
Rejected $\nu_\mu$ CC events typically have higher inelasticity and dimmer muon tracks, which reduce the track fit quality.
Approximately 65\% of all other interactions (i.e. $\nu_{e,\tau}$ CC and all NC) are rejected, leading to approximately 70\% purity of $\nu_\mu$ CC interactions at the final level\footnote{The signal purity is estimated at the best-fit point of the analysis}.
The reconstructed muon direction $\theta_{z, reco}$ is used as an estimate for the arrival direction of the interacting neutrino.
The zenith angle of the muon is calculated from the fitted tracklike hyperbolic pattern. The median neutrino zenith resolution is approximately 12$^{\circ}$ at 10 GeV and improves to 6$^{\circ}$ at 40 GeV. 

The neutrino energy reconstruction assumes the existence of a muon track and a hadronic shower at the neutrino interaction point.  Muons selected for this analysis are in the minimum ionizing regime~\cite{MuonLosses_GROOM2001183}. The energy of these muons is, therefore, determined by their range $R_{\mu}$. The total neutrino energy is then calculated as the sum of the energies attributed to the hadronic shower ($E_{shower}$) and the muon track,
\begin{equation}
\label{totalrecoenergy}
E_{reco} \approx E_{shower} + a R_{\mu},
\end{equation} 
where \unit[$a \approx 0.23 $]{GeV/m} is the constant\footnote{An additional term is used in the energy reconstruction to account for the rising muon losses at higher energies. However, its impact is small and therefore is not shown in Eq. (\ref{totalrecoenergy}) } energy loss of muons in ice. The muon range is calculated by identifying the starting and stopping points of a muon along the reconstructed track direction.  
The energy reconstruction is described in more detail in \cite{DeepCore_PhysRevD.91.072004}. The median energy resolution is about 30\% at 8 GeV and improves to 20\% at 20 GeV.

\begin{figure}[!ht]
  \centering
    \includegraphics[width=0.5\textwidth]{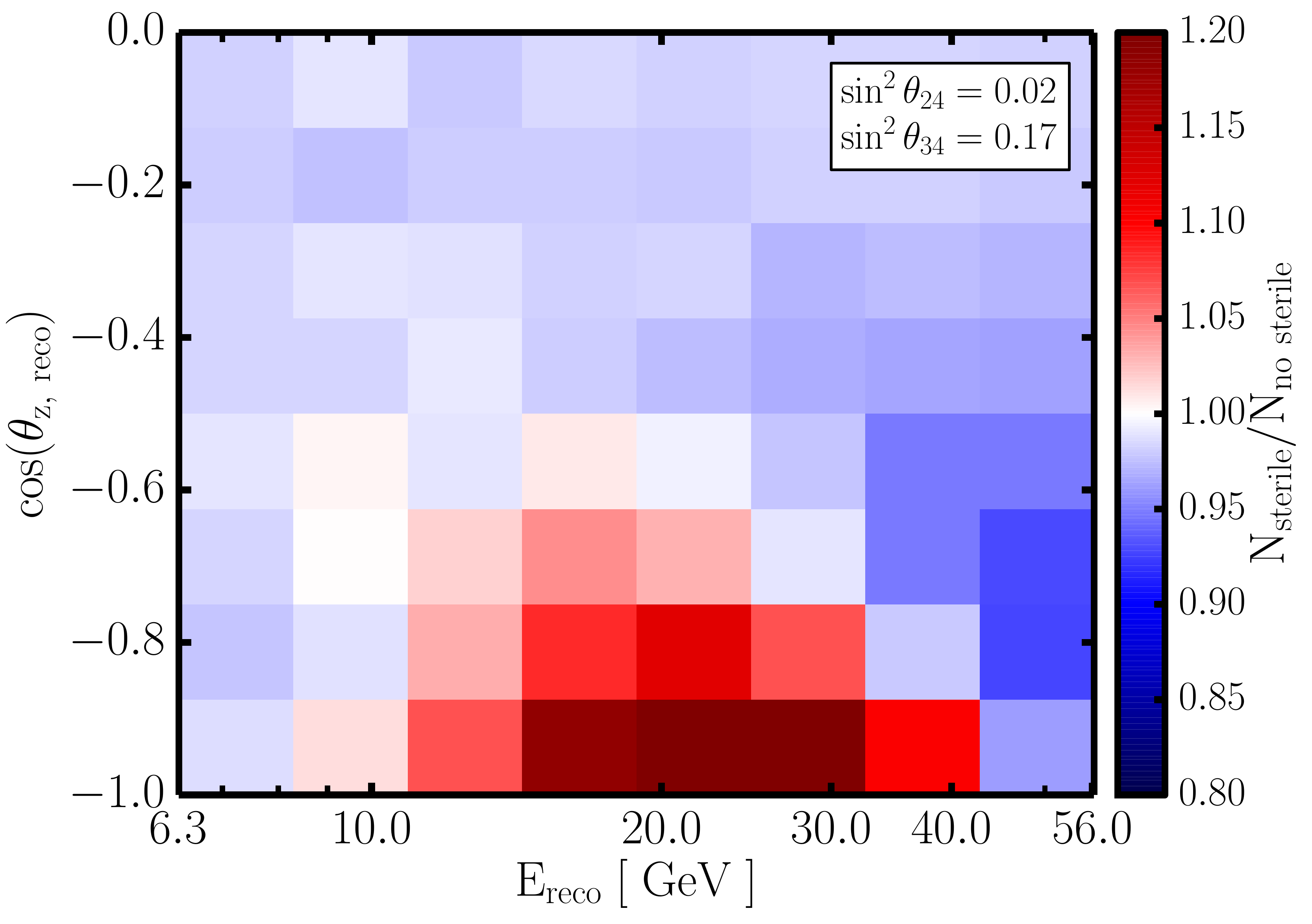}
  \caption{\label{signature} The ratio of the expected event counts for a sterile neutrino hypothesis and the case of no sterile neutrino. Sterile neutrino mixing parameters  $\sin^2 \theta_{24} = 0.02$ and $\sin^2 \theta_{34} = 0.17$ are assumed. The values $\Delta m_{32}^2 = 2.52 \cdot 10^{-3} $ eV$^2$ and $\sin^2 \theta_{23} = 0.51$ are assumed for the standard atmospheric mixing parameters. Both expectations are normalized to the same  total number of events. }
\end{figure}

\section{Data analysis techniques}
\label{sec:data_analysis}

Three years of DeepCore data~\cite{IC86datarelease}, comprising 5118 events at the final level, are used in this study. They are compared to predictions from simulations as described in the following subsections. 

\subsection{Monte Carlo simulation}

Neutrino interactions and hadronization processes are simulated using GENIE~\cite{GENIE_Andreopoulos:2009rq}. Produced muons are propagated with PROPOSAL~\cite{PROPOSAL_Koehne20132070}. GEANT4 is used to propagate hadrons and particles producing  electromagnetic showers with energies less than 30 GeV and 100 MeV, respectively. Light output templates~\cite{Leif_thesis} are used for particles with higher energies. Clsim~\cite{CLSim} is used to propagate the resulting photons. 
The equivalent of 30 years of detector operation is simulated for each neutrino flavor. This ensures that the Poisson fluctuations due to Monte Carlo statistics are much smaller than statistical uncertainties in the data and, therefore, can be neglected throughout the analysis. 

\subsection{Signal signature}
\begin{table*}
\caption{\label{tab:nuisance}%
The physics parameters of interest and their best-fit points obtained in the analysis for normal (NO) and inverted (IO) neutrino mass orderings are shown. The nuisance parameters used to account for systematic uncertainties, their priors (if used) and their best-fit values are also given. 
}
\begin{ruledtabular}
\begin{tabular}{lccc}
\textrm{ \textbf{Parameter}}&
\textrm{ \textbf{Priors} }& 
\textrm{ \textbf{Best fit (NO)}} &
\textrm{ \textbf{Best fit (IO)}}\\
    \colrule
    \multicolumn{4}{c}{\textbf{Sterile mixing parameters}}\\
    \colrule
$|U_{\mu4}|^2$ & no prior & 0.00 & 0.00\\
$|U_{\tau4}|^2$ & no prior & 0.08 & 0.06 \\
    \colrule
    \multicolumn{4}{c}{\textbf{Standard mixing parameters}}\\
    \colrule
$\mathrm{\Delta m_{32}^2\ [\ 10^{-3} eV^2\ ]}$      &    no prior   & 2.52 & $\mathrm{- 2.61}$\\
$ \sin^2 \theta_{23}$    &    no prior & $\mathrm{0.541}$   & $\mathrm{0.473} $                                     \\
    \colrule
    \multicolumn{4}{c}{\textbf{Flux parameters}}\\
    \colrule
$\gamma$          & no prior            &   --2.55 &     --2.55 \\
$\nu_e$ normalization    & $\mathrm{1\pm0.05}$ & 0.996 &         0.997 \\
$\Delta (\nu / \bar{\nu})$, energy dependent & $\mathrm{0\pm1\sigma}$ &  $0.19\mathrm{\sigma}$ &    $0.21\mathrm{\sigma}$ \\
$\Delta (\nu / \bar{\nu})$, zenith dependent & $\mathrm{0\pm1\sigma}$ &  $0.19\mathrm{\sigma}$ &    $0.16\mathrm{\sigma}$ \\
    \colrule
    \multicolumn{4}{c}{\textbf{Cross section parameters}}\\
    \colrule
$M_A$ (resonance) [ GeV ]  & $\mathrm{1.12\pm0.22}$& 1.16 & 1.14 \\
$M_A$ (quasielastic)  [ GeV ]   & $\mathrm{0.99^{{}+0.25}_{{}-0.15}}$ & 1.03 &  1.03 \\
    \colrule
    \multicolumn{4}{c}{\textbf{Detector parameters}}\\
    \colrule
Hole ice scattering {$\mathrm{[\ cm^{-1}}$\ ]} & $\mathrm{0.02\pm0.01}$ & $\mathrm{0.021}$ & $\mathrm{0.021}$ \\
DOM efficiency $\mathrm{[\ \%\ ]}$&$\mathrm{100\pm10}$  & 101 &  101 \\
    \colrule
    \multicolumn{4}{c}{\textbf{Background}}\\
    \colrule
Atm. $\mu$ contamination $\mathrm{[\ \%\ ]}$ & no prior& 0.01 & 0.4 \\
\end{tabular}
\end{ruledtabular}
\end{table*}

The impact of a sterile neutrino on the event rate as a function of reconstructed energy and zenith in this study is shown in Fig.~\ref{signature}. 
The most dramatic changes are expected at reconstructed energies between 20 and 30 GeV for neutrinos crossing the Earth's core ($\cos \theta_z \lesssim -0.85$). In addition, the presence of a sterile neutrino changes the normalization as described in Sec.~\ref{sec:mixing}. This gives an approximately uniform deficit of events seen in other regions of reconstructed energy and zenith.

\subsection{Fitting procedure}

A binned maximum log-likelihood algorithm with nuisance parameters \cite{PDG_1674-1137-38-9-090001} to account for systematic uncertainties is used to determine the sterile neutrino mixing parameters. The data are binned in an 8$\times$8 histogram in $\cos{\theta_{z,reco}}$ and $\log E_{reco}$. Only tracklike events with $\cos \theta_{z, reco} \in [ -1, 0 ] $ and $E_{reco} \in [ 10^{0.8}, 10^{1.75} ]$ GeV are used in the analysis. The log-likelihood is defined as 
\begin{equation}
\label{LLH}
- \ln \mathcal{L}  = \sum_{i}(\mu_i -  n_i \ln \mu_i) + \sum_k^{n_{priors}} \frac{(\phi_k - \phi_k^{0})^2}{2 \sigma_{\phi_k}^2},
\end{equation}
where $n_i$ is the number of events in the $i$th bin of a data histogram,  and $\mu_i=\mu_i (\bar{\theta}, \bar{\phi})$ is the expected number of events from the physics parameters $\bar{\theta}$ and nuisance parameters $\bar{\phi}$. The second term of Eq.~(\ref{LLH}) accounts for the prior knowledge of the nuisance parameters, where $\phi^0_{k}$ and $\sigma_{\phi_k}$ are the estimated value and uncertainty, respectively, on the parameter $\phi_k$. The priors come from independent measurements or uncertainties in model predictions. 
As stated in Sec.~\ref{sec:mixing}, the physics parameters of interest for this study are the mixing angles $\theta_{24}$ and $\theta_{34}$. Confidence levels are estimated using Wilks's theorem \cite{wilks1938} for the difference $-2 \Delta \ln \mathcal{L} $ between the profile log-likelihood  and the log-likelihood at the best-fit point.

The expected histogram bin content is obtained by event-by-event re-weighting of events in Monte Carlo simulations. In addition, the impact of the detector systematic uncertainties is estimated at the histogram level. 

\subsection{Treatment of systematic uncertainties}

Eleven nuisance parameters, listed in Table \ref{tab:nuisance}, are used in the analysis to account for the impact of systematic uncertainties in this study. These systematic uncertainties are grouped in five classes and are  explained in the following sections.

\subsubsection{Neutrino mixing}

The values of the standard atmospheric mixing parameters determine the neutrino oscillations pattern. The value of the mass splitting $\Delta m_{32}^2$ defines the position of the minimum and  $\theta_{23}$ is related to its amplitude. Similar modifications of the oscillations pattern, but limited to the neutrinos crossing the Earth's core, are caused by the addition of a sterile neutrino. This makes standard mixing parameters the most important uncertainties for this study.

\begin{figure*}[!ht]
  \centering
    \includegraphics[width=0.9\textwidth]{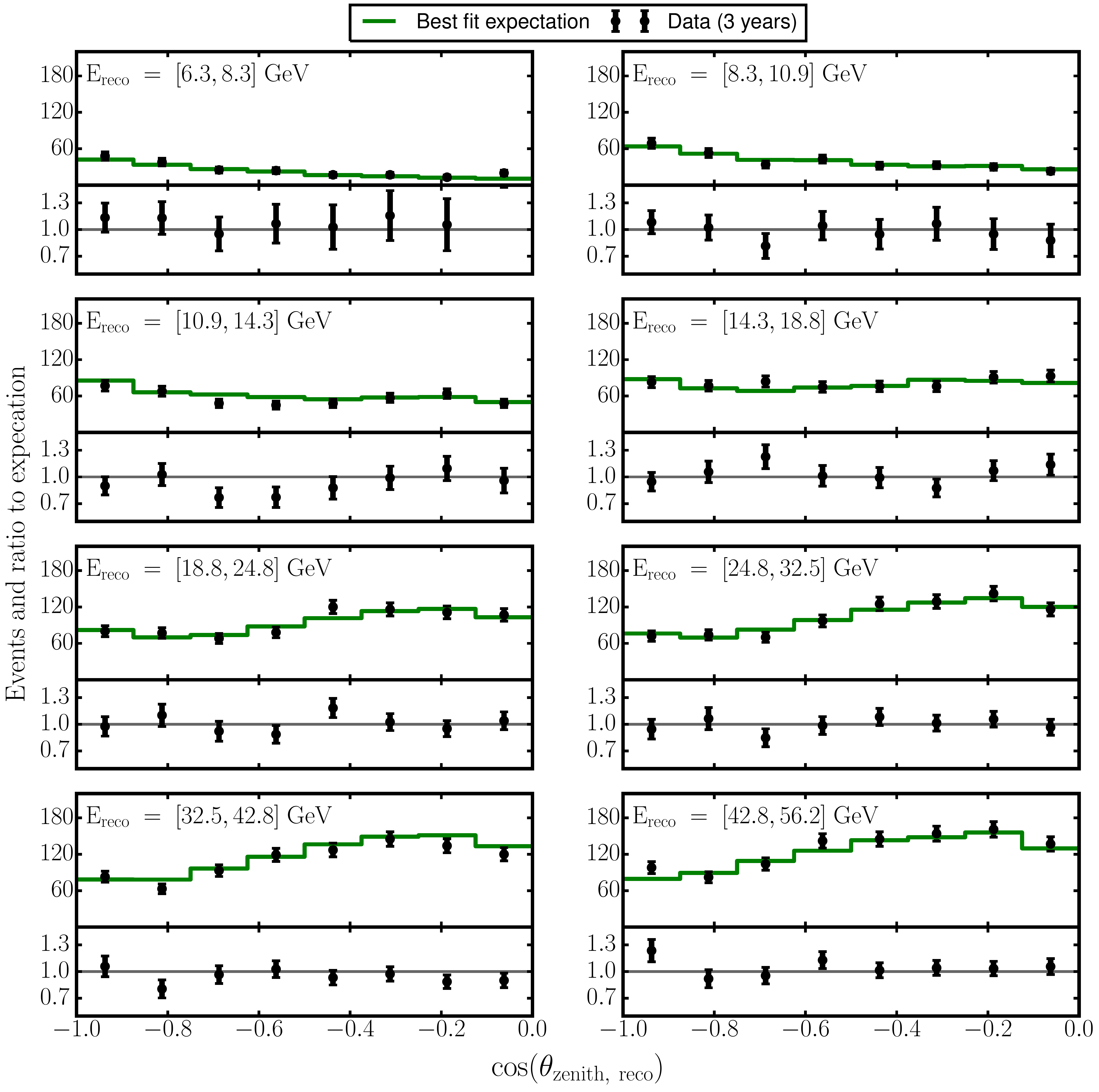}
  \caption{  \label{fig:dataMCbins} The comparison of the data (black dots) and the expectation at the best-fit point for the bins used in the analysis. The expectation at the best fit includes a full calculation of the oscillation probabilities for the ``3+1" model, impact of systematic uncertainties and background.}
\end{figure*} 

\begin{figure}[!ht]
  \centering
    \includegraphics[width=0.5\textwidth]{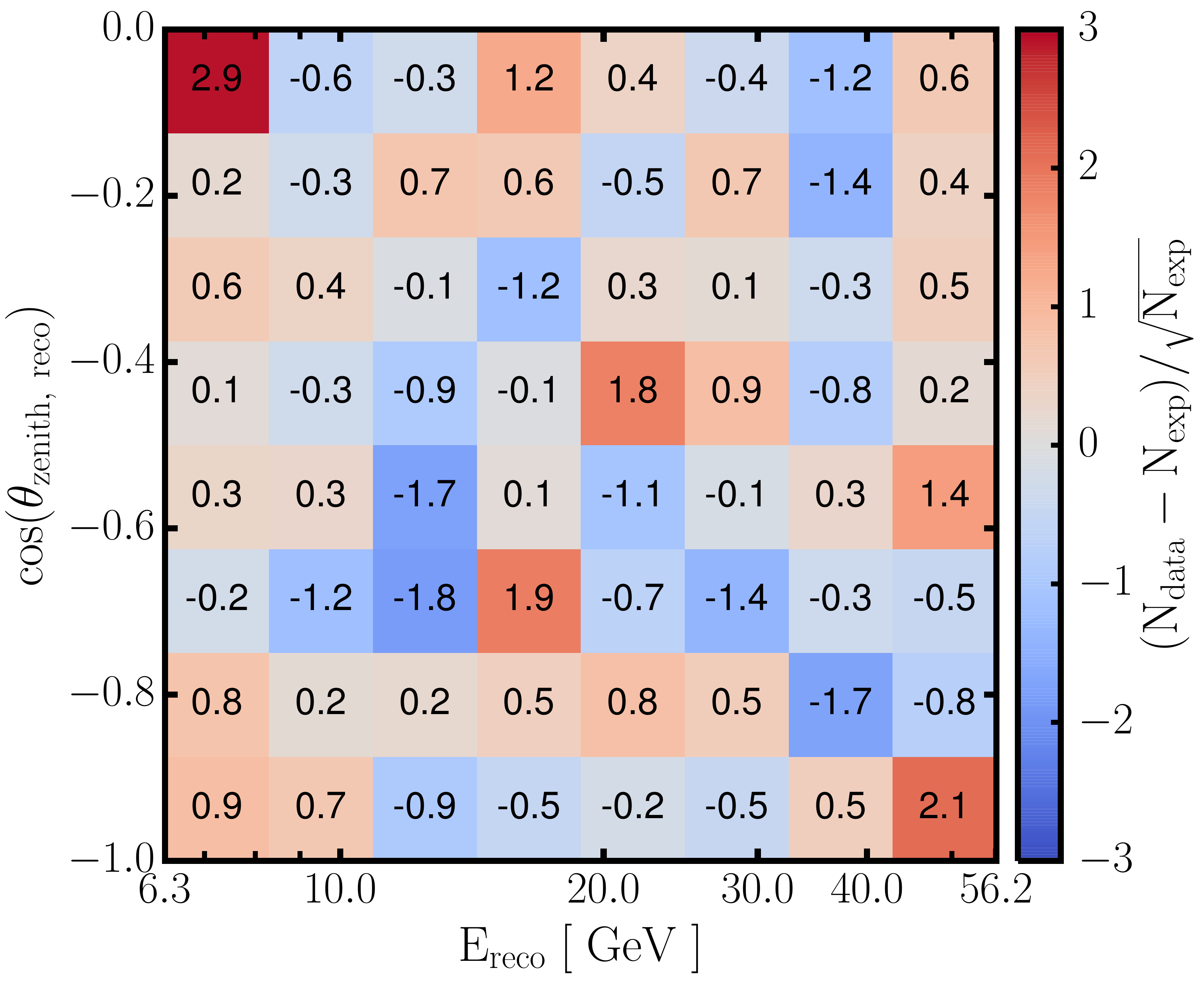}
  \caption{ \label{fig:pulls} Statistical pulls between data and expectation for the best-fit point.  }
\end{figure}

Simulations show that prior values for the standard mixing parameters can lead to a fake nonzero best-fit point with significance on the order of 1 $\sigma$. Also, the global values of $\Delta m_{32}^2$ and $\theta_{23}$  do not include sterile neutrinos in the model.  Therefore, no priors on the standard mixing parameters are used in this study. 
Values of other mixing parameters such as $\theta_{12}$, $\theta_{13}$, and $\Delta m_{21}^2$ are found to have no impact on the analysis and are fixed to the global best-fit values from~\cite{PDG_1674-1137-38-9-090001}. Both normal  ($m_1 < m_2 < m_3 < m_4$) and inverted ($ m_2 < m_3 < m_1 < m_4$) neutrino mass orderings are considered in the analysis.

\begin{figure}[!ht]
  \centering
    \includegraphics[width=0.5\textwidth]{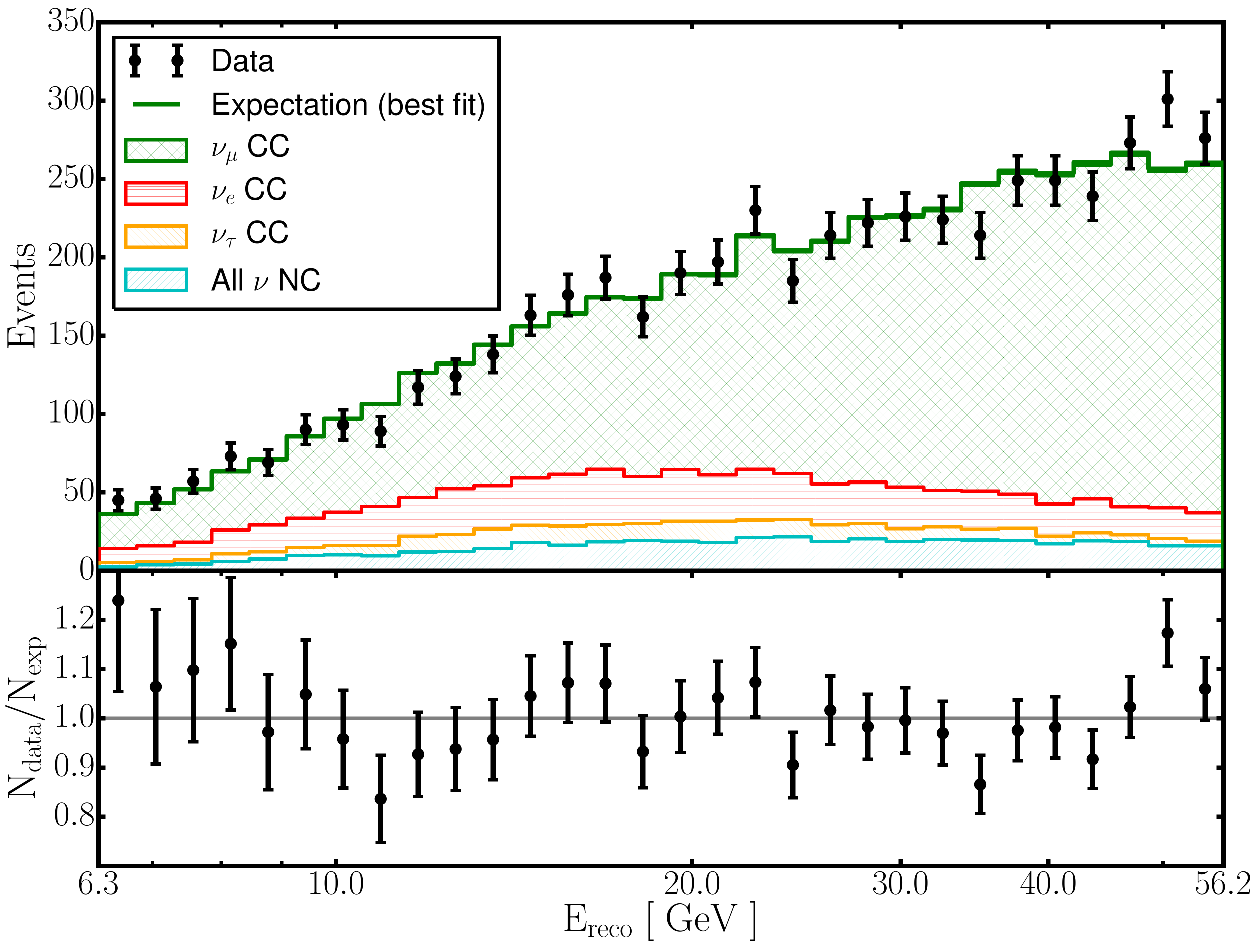}
    \includegraphics[width=0.5\textwidth]{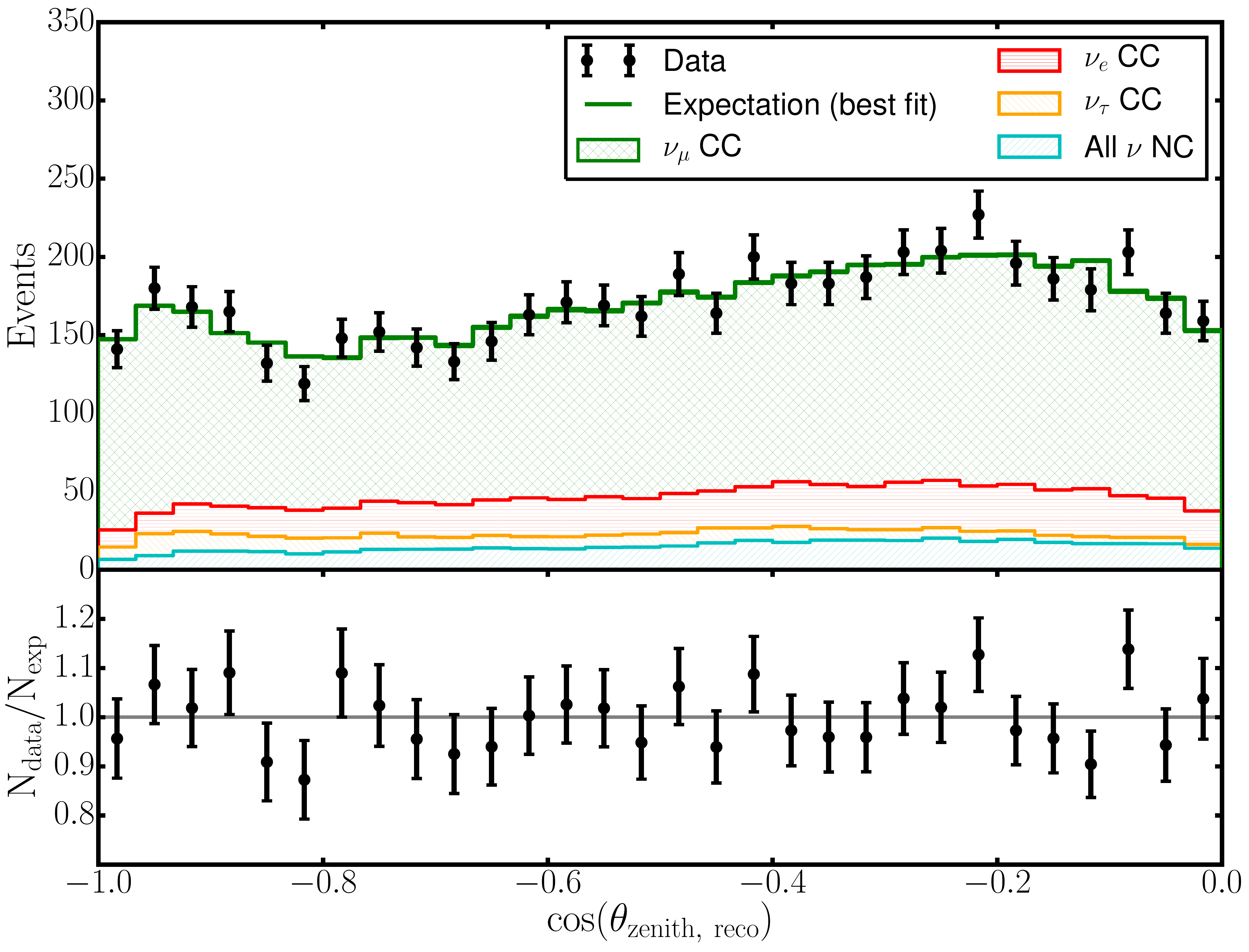}
 \caption{ \label{fig:projections} Event rates shown as a function of (top) $E_{reco}$ and (bottom) $\cos \theta_{z,reco}$. The various different neutrino components from Monte Carlo simulation used in the fit are shown as stacked histograms. The total expected event rate is in good agreement with the observed data, shown as black points.}
\end{figure}

\subsubsection{Flux systematics}
The neutrino flux model from \cite{Honda2015_PhysRevD.92.023004}, which  assumes a nominal value of $\gamma = -2.66$ for the cosmic ray spectral index, is used in the analysis.
The effects of several systematic uncertainties, such as the properties of the global ice model and deep inelastic scattering cross section, are degenerate with a change in the spectral index. Therefore, this nuisance parameter is left unconstrained in the fit to account for these subdominant uncertainties.

The normalization of the $\nu_e$ flux is assigned a 5\% Gaussian prior. The uncertainties of the neutrino and antineutrinos fractions of the neutrino flux from \cite{Barr:2006it} are used. Their deviations from the flux model are parametrized as two independent parameters describing energy dependent $\Delta(\nu/\bar{\nu})_{energy}$ and zenith dependent $\Delta(\nu/\bar{\nu})_{zenith}$ uncertainties. The overall normalization of the flux is left unconstrained to account for large uncertainties on the absolute flux of atmospheric neutrinos. 
\subsubsection{Cross section systematics}
The main interaction process for neutrinos in the energy range of this analysis is deep inelastic scattering (DIS). Uncertainties of the DIS cross sections are taken into account as modifications of an effective spectral index and the overall normalization of the flux. Uncertainties of non-DIS processes, such as resonant and quasielastic scattering, are estimated by GENIE as a correction to the weights of the generated interactions. 
This is done by varying the axial mass form factors $M_A$ as described in~\cite{GENIE_Andreopoulos:2015wxa}.
\subsubsection{Detector systematics}

Uncertainties on the detector properties, like the efficiency of the optical modules and their angular acceptance, have a large impact in this analysis.

To estimate the impact of the DOM efficiency, seven discrete Monte Carlo sets are used. They span the  range of 85--115\% of the nominal efficiency in steps of 5\%. 
Each set is processed using the event selection described in Sec.~\ref{sec:selection} and the final events are binned in reconstructed energy and $\cos \theta_{z,reco}$ to produce expectation histograms analogous to Fig.~\ref{signature}. The impact of varying the efficiency continuously is then estimated by fitting a second degree polynomial to the changing event rate obtained from the discrete sets in each analysis bin.
A Gaussian prior centered at the nominal efficiency (100\%) with a $\sigma$ of 10 \% is applied. 

One of the most important systematic uncertainties is the DOM angular acceptance. During the deployment of  IceCube strings holes were drilled into the ice with a hot water drill. After the refreezing process, the ice along the strings has different optical properties in comparison to other part of the detector. This process effectively changes the angular acceptance of DOMs. Its impact is especially important for the low energy neutrinos in DeepCore, because such events leave only a small signal in the detector. The properties of the refrozen ice, such as effective scattering length, change the angular profile of reconstructed events. This systematic uncertainty is treated in a similar way to the DOM efficiency. Ten discrete systematic sets with different effective scattering coefficients between 0.01 \unit[]{$\mathrm{cm^{-1}}$} and 0.033 \unit[]{$\mathrm{cm^{-1}}$} are used to determine a bin-by-bin effect of the refrozen ice properties on the event rate. The effect for the intermediate values is estimated using third degree polynomials.   
A Gaussian prior of $0.02\pm0.01\mathrm{~cm^{-1}}$ is applied.

\subsubsection{Background}

It is also important to estimate the impact of the background due to atmospheric muons reaching DeepCore. 
The rejection algorithms for atmospheric muons are developed using Monte Carlo simulations produced with CORSIKA \cite{CORSIKA_Heck:1998vt}. However, producing enough muon statistics at the final analysis level is computationally intensive and cannot be performed with currently available resources.  Therefore, the impact of the muon background is addressed using the data-driven template explained in Sec.~\ref{sec:selection}. The muon template is then added to the expected event rate from neutrino events to form a total expectation. 
Its normalization is left unconstrained to assess the impact from the atmospheric muon background. The selection of direct photons successfully removes events from pure electronic noise, and, therefore, such noise is not considered in this study.

\section{Results}
\label{sec:results}

The data are found to be consistent with the standard three-neutrino hypothesis. 
Predictions from neutrino simulations and the atmospheric muon template fit the experimental data well with  a $\chi^2$  of 54.9. 
There are 64 data bins in total fitted with 13 parameters. 
Some of the parameters effectively contribute less than one degree of freedom (d.o.f) due to priors and correlations. 
The number of d.o.f. is estimated by fitting 2000 statistical trials obtained by fluctuating the expectation from the detector simulations and background. 
This exercise provides a goodness of fit distribution that is then fit with a $\chi^2$ distribution to extract the effective number of d.o.f.
The resulting number of d.o.f. is estimated to be $56.3\pm0.3$ and the probability to obtain the observed $\chi^2$ is, therefore, 53\%. 

The agreement between the data and the expectation at the best-fit point is shown in Fig.~\ref{fig:dataMCbins} for the bins used in the fit. The bin-by-bin pulls of the data compared to the expectation at the best-fit point are shown in Fig.~\ref{fig:pulls}. The pulls are distributed in the way expected from statistical fluctuations without large deviations or clustering in specific energy or zenith ranges.

\begin{figure}[!ht]
  \centering
    \includegraphics[width=0.5\textwidth]{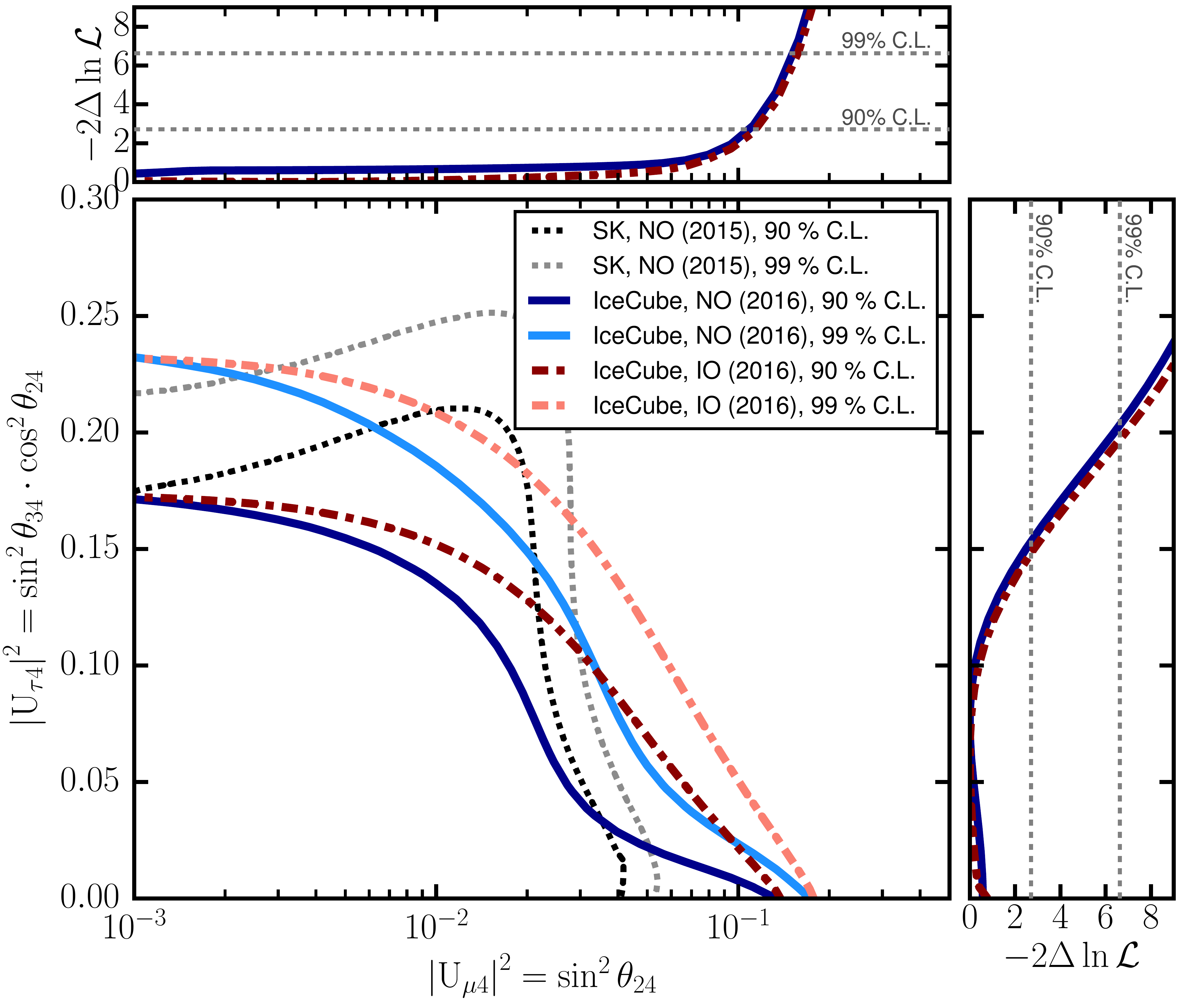}
  \caption{\label{fig:contours} 
The results of the likelihood scan performed in the analysis. 
The solid lines in the larger panel show the exclusion limits set in this study at 90-\% (dark blue) and 99-\% C.L. (light blue) assuming the normal neutrino (NO) mass ordering and using critical values from $\chi^2$ with 2 d.o.f. 
The dark (light) red dash-dotted lines represent the 90-\% (99-\%) C.L. exclusions assuming an inverted mass ordering (IO).
The dashed lines show the exclusion from the Super-Kamiokande experiment \cite{SuperK_sterile_PhysRevD.91.052019}. 
The top and right panels show the projection of the likelihood on the mixing matrix elements $|U_{\mu4}|^2$ and  $|U_{\tau4}|^2$, respectively.  }
\end{figure}

The upper and lower parts of Fig.~\ref{fig:projections} depict distributions of $E_{reco}$ and $\cos \theta_{z, reco}$, respectively. 
It also shows the expectation from the different components of the simulations used in the fit. The dominant contribution comes from $\nu_\mu$ CC interactions with some contamination from $\nu_e$, $\nu_\tau$ and NC interactions of all flavors.  
The atmospheric muon contamination is fit to about 0.4 \% and, therefore, not shown in Fig.~\ref{fig:projections}. 

All nuisance parameters are fit near the nominal values; their values can be found in Table \ref{tab:nuisance}. Inverted mass ordering is marginally preferred in the fit. 
The best estimates of the sterile mixing parameters are given in Table \ref{tab:nuisance}.
The difference between the best fit and the standard three-neutrino hypothesis is $-2 \Delta \ln \mathcal{L} = 0.8$. Such a value is expected from statistical fluctuations of the data with 30\% probability estimated from the aforementioned 2000 trials.

Exclusion contours are obtained by scanning the likelihood space in $\left | U_{\mu4}\right |^2$ vs $\left | U_{\tau4}\right |^2$ and are presented in Fig.~\ref{fig:contours}. The corresponding limits on the elements of the mixing matrix are
\begin{equation}
\begin{split}
\left |U_{\mu 4} \right |^2 < 0.11\ \mathrm{(90\%\ C.L.)}, 
\\ 
\left |U_{\tau 4} \right |^2 < 0.15\ \mathrm{(90\%\ C.L.)}, 
\end{split}
\label{eq:limits}
\end{equation}
where the confidence levels are obtained using Wilks's theorem. 

The best-fit values for the standard neutrino mixing parameters  are \unit[$\Delta m_{32}^2 = 2.52 \cdot 10^{-3}$ ]{$\mathrm{eV^{2}}$} and \unit[$\sin^2 \theta_{23} = 0.541$]{} (assuming normal neutrino mass ordering), which are different from the results of~\cite{DeepCore_3yrs_disappearance_PhysRevD.91.072004}. 
The best-fit point for $\Delta m^2_{32}$ is now 1 $\sigma$ lower compared to the previous measurement.  Although the data set and analysis methods used in the two analyses are similar, there are a few differences responsible for the change. Since the publication of~\cite{DeepCore_3yrs_disappearance_PhysRevD.91.072004} the Monte Carlo simulation and event reconstruction have been improved.  In particular, there is a new charge calibration used for the PMTs in simulation that leads to an update of the effective energy scale in the detector reconstruction. This leads to a change in the reconstructed position of the muon disappearance minimum, which is proportional to $\Delta m_{32}^2$. A more stringent event selection is also implemented to improve atmospheric muon background rejection; however, the impact on the measurement of the atmospheric mixing angle is small ($<$0.3 $\sigma$).

\section{Conclusions and outlook}
\label{sec:conclusions}

Figure~\ref{fig:contours} shows the exclusion contours obtained in this study compared to a search performed by the Super-Kamiokande experiment~\cite{SuperK_sterile_PhysRevD.91.052019}, where the limit $\left | U_{\tau4}\right |^2 < 0.18$ (90~\% C.L.) is obtained.  
Using three years of IceCube DeepCore data improves the world best limit on the $\left | U_{\tau4}\right |^2$ element by approximately 20~\% at 90~\% C.L. 
The MINOS experiment also derives a constraint on $\left | U_{\tau4}\right |^2 < 0.20$ (90~\% C.L.)~\cite{MINOS_sterile_PhysRevLett.117.151803}, however this limit is only provided for a single mass splitting of $\Delta m^2_{41}=0.5$~eV${}^2$. 
As there is no explanation of how that result scales with $\Delta m^2_{41}$, it is difficult to compare with the results obtained with IceCube DeepCore.

The best constraints on $|U_{\mu4}|^2$ come from the IceCube study using TeV neutrinos~\cite{IC86_HEsteriles_PhysRevLett.117.071801} and the MINOS experiment~\cite{MINOS_sterile_PhysRevLett.117.151803}. The sensitity of this study to $|U_{\mu4}|^2$  is limited by a number of factors, including flux uncertainties and detector resolutions, that result in a degeneracy with other parameters of the analysis.

Current global fits of the neutrino oscillations experimental data suggest $\left | U_{e4}\right |^2 = 0.023-0.028$, where the range covers values presented in~\cite{Kopp2013,Global_Collin:2016aqd,Global_Capozzi:2016vac,Global_Collin:2016rao}.
In this study $\left | U_{e4}\right |^2$ is assumed to be zero. The impact of a possible nonzero value is estimated by fitting $\theta_{14}$ as a nuisance parameter with prior approximately 4 times larger than the current global fit estimate. This prior accounts for both zero and nonzero values of $\theta_{14}$. 
Because of the relatively small $\nu_e$ contamination of the data sample, the value of $\left | U_{e4} \right |^2$ allowed by the current global fits has no impact on the analysis.

The value of $\Delta m_{41}^2$ was fixed at \unit[1.0]{$\mathrm{ eV^2}$} throughout this analysis. Changing the value of $\Delta m_{41}^2$ in the range between \unit[0.1 and 10.0 ]{$\mathrm{eV^2}$} has no impact on the limit on $\left | U_{\tau4} \right |^2$. 
The limit on $\left | U_{\mu4} \right |^2$ depends only weakly on $\Delta m_{41}^2$. At \unit[0.1]{$\mathrm{eV^2}$} it degrades to 0.12, representing an \unit[8]{\%} relative change in the exclusion limit, while at \unit[10]{$\mathrm{eV^2}$} we observe a relative improvement in the limit by \unit[9]{\%}. 

Monte Carlo studies show that the current limits on the sterile neutrino mixing are statistically limited and can be improved using more data collected by IceCube DeepCore. Extending the energy range may yield more information about the flux and its normalization and thus better constrain  systematic uncertainties. 
Furthermore, inclusion of cascadelike events may open a possibility to use the $\nu_e$ and $\nu_\tau$ components of the flux and NC interactions to improve the sensitivity to the sterile neutrino mixing. 

\begin{acknowledgments}

We acknowledge the support from the following agencies:
U.S. National Science Foundation-Office of Polar Programs,
U.S. National Science Foundation-Physics Division,
University of Wisconsin Alumni Research Foundation,
the Grid Laboratory Of Wisconsin (GLOW) grid infrastructure at the University of Wisconsin - Madison, the Open Science Grid (OSG) grid infrastructure;
U.S. Department of Energy, and National Energy Research Scientific Computing Center,
the Louisiana Optical Network Initiative (LONI) grid computing resources;
Natural Sciences and Engineering Research Council of Canada,
WestGrid and Compute/Calcul Canada;
Swedish Research Council,
Swedish Polar Research Secretariat,
Swedish National Infrastructure for Computing (SNIC),
and Knut and Alice Wallenberg Foundation, Sweden;
German Ministry for Education and Research (BMBF),
Deutsche Forschungsgemeinschaft (DFG),
Helmholtz Alliance for Astroparticle Physics (HAP),
Research Department of Plasmas with Complex Interactions (Bochum), Germany;
Fund for Scientific Research (FNRS-FWO),
FWO Odysseus programme,
Flanders Institute to encourage scientific and technological research in industry (IWT),
Belgian Federal Science Policy Office (Belspo);
University of Oxford, United Kingdom;
Marsden Fund, New Zealand;
Australian Research Council;
Japan Society for Promotion of Science (JSPS);
the Swiss National Science Foundation (SNSF), Switzerland;
National Research Foundation of Korea (NRF);
Villum Fonden, Danish National Research Foundation (DNRF), Denmark

\end{acknowledgments}

\interlinepenalty=10000

\end{document}